\documentclass[aps,prl,floatfix,twocolumn,nofootinbib,preprintnumbers]{revtex4}

\usepackage[english]{babel}
\usepackage{amsfonts,amsmath,amssymb,amsbsy,amsthm,bm,mathtools,fixmath}
\usepackage{comment,enumerate,footnote,graphicx,subfloat,relsize,footnote}
\usepackage{array,tabularx,tabu,multirow,framed}
\usepackage[utf8]{inputenc}

\usepackage{tikz}
\usetikzlibrary{arrows,shapes}
\usetikzlibrary{trees,patterns}
\usetikzlibrary{matrix,arrows} 				
\usetikzlibrary{positioning}				  
\usetikzlibrary{calc,through}				  
\usetikzlibrary{decorations.pathreplacing}  
\usepackage{pgffor}							

\usetikzlibrary{decorations.pathmorphing}	
\usetikzlibrary{decorations.markings}
\tikzset{
	>=stealth', 
    vector/.style={decorate, decoration={snake}, draw},
	provector/.style={decorate, decoration={snake,amplitude=2.5pt}, draw},
	antivector/.style={decorate, decoration={snake,amplitude=-2.5pt}, draw},
	bigvector/.style={decorate, decoration={snake,amplitude=4pt}, draw},
    fermion/.style={draw=black, postaction={decorate},
        decoration={markings,mark=at position .55 with {\arrow[draw=black]{>}}}},
    fermionbar/.style={draw=black, postaction={decorate},
        decoration={markings,mark=at position .55 with {\arrow[draw=black]{<}}}},
    fermionnoarrow/.style={draw=black},
    gluon/.style={decorate, draw=black,
        decoration={coil,amplitude=4pt, segment length=5pt}},
    scalar/.style={dashed,draw=black, postaction={decorate},
        decoration={markings,mark=at position .55 with {\arrow[draw=black]{>}}}},
    scalarbar/.style={dashed,draw=black, postaction={decorate},
        decoration={markings,mark=at position .55 with {\arrow[draw=black]{<}}}},
    scalarnoarrow/.style={dashed,draw=black},
    momentum/.style={draw=black, postaction={decorate},
        decoration={markings,mark=at position 1 with {\arrow[draw=black]{>}}}},
    antimomentum/.style={draw=black, postaction={decorate},
        decoration={markings,mark=at position 0.1 with {\arrow[draw=black]{<}}}}
}

\tikzstyle{block} = [draw, rectangle, 
    minimum height=3em, minimum width=6em]
    
\newcommand{\nc}{\newcommand}
\nc{\pd}{\partial}
\nc{\bea}{\begin{eqnarray}}
\nc{\eea}{\end{eqnarray}}
\nc{\bal}{\begin{alignedat}}
\nc{\eal}{\end{alignedat}}
\nc{\beq}{\begin{equation}}
\nc{\eeq}{\end{equation}}
\nc{\bit}{\begin{itemize}}
\nc{\eit}{\end{itemize}}
\nc{\benu}{\begin{enumerate}}
\nc{\eenu}{\end{enumerate}}
\nc{\bdes}{\begin{description}}
\nc{\edes}{\end{description}}
\nc{\bma}{\begin{pmatrix}}
\nc{\ema}{\end{pmatrix}}

\nc{\nn}{\nonumber}
\nc{\hc}{\text{h.c.}}
\nc{\cc}{\text{c.c.}}

\nc{\slashed}[1]{{#1}\hspace{-2mm}/}

\nc{\abs}[1]{\left| #1 \right|}
\def\[{\left[}
\def\]{\right]}
\def\({\left(}
\def\){\right)}
\def\<{\langle}
\def\>{\rangle}

\def\g5{\gamma_{5}}

\def\GeV{{\rm GeV}}
\def\TeV{{\rm TeV}}

\def\g{\gamma}
\def\d{\delta}

\def\m{\mu}

\def\s{\sigma}

\def\f{\phi}
\def\x{\chi}

\def\ag		{\alpha_g}
\def\ah		{\alpha_h}

\def\as			{\alpha_s}

\def\DM			{\mathsmaller{\rm DM}}
\def\eq			{{\rm eq}}

\def\gstareffsqrt	{g_{*,\rm eff}^{1/2}}
\def\gh			{g_h}
\def\gs			{g_s}
\def\gstarS		{g_{*\mathsmaller{S}}}
\def\gx			{g_\x}

\def\mh			{m_h}

\def\mx			{m_\x}
\def\mX			{m_{\mathsmaller{X}}}
\def\sigmaeff	{\sigma_{\rm eff}}
\def\Xdagger	{X^\dagger}

\def\YX			{Y_{X}}
\def\YXdagger	{Y_{\Xdagger}}
\def\vrel		{v_{\rm rel}}
\def\zetag		{\zeta_g}
\def\zetah		{\zeta_h}

\renewcommand{\vec}{\textbf}

\newcommand{\AddrLPTHE}{Laboratoire de Physique Th\'eorique et Hautes Energies (LPTHE), UMR 7589 CNRS \& UPMC, 4~Place Jussieu, F-75252, Paris, France}
\newcommand{\AddrILP}{Sorbonne Universit\'es, Institut Lagrange de Paris (ILP), 98 bis Boulevard Arago, 75014 Paris, France}
\newcommand{\AddrNIKHEF}{Nikhef, Science Park 105, 1098 XG Amsterdam, The Netherlands}

\begin{document}

\title{Higgs Enhancement for the Dark Matter Relic Density}

\author{Julia Harz}
\email{jharz@lpthe.jussieu.fr}
\affiliation{\AddrILP}
\affiliation{\AddrLPTHE}

\author{Kalliopi Petraki} 
\email{kpetraki@lpthe.jussieu.fr}
\affiliation{\AddrLPTHE}
\affiliation{\AddrNIKHEF}

\preprint{Nikhef-2017-058}

\begin{abstract}

We consider the long-range effect of the Higgs on the density of thermal-relic dark matter. While the electroweak gauge boson and gluon exchange have been previously studied, the Higgs is typically thought to mediate only contact interactions. We show that the Sommerfeld enhancement due to a 125 GeV Higgs can deplete TeV-scale dark matter significantly, and describe how the interplay between the Higgs and other mediators influences this effect. We discuss the importance of the Higgs enhancement in the Minimal Supersymmetric Standard Model, and its implications for experiments.

\end{abstract}

\maketitle

\section{Introduction}

The dark matter (DM) density has been determined to an unprecedented precision by the \emph{Planck} satellite \cite{Ade:2015xua}
\beq
\Omega_{\DM} h^2 = 0.1199 \pm 0.0022 \,.
\label{eq:Omega}
\eeq
This measurement provides a powerful constraint on DM theories. However, in order to constrain DM models reliably, it is essential to compute the expected DM density in a comprehensive manner.

It is well known that the density of thermal-relic DM is determined by the strength of the DM annihilation processes. These include the DM self-annihilations, as well as the co-annihilations of DM with particles of similar mass. The latter were first discussed in~\cite{Edsjo:1997bg}, but have received renewed attention recently, in particular in the context of  DM coupled to the Weak Interactions of the Standard Model (SM), known as WIMP DM~\cite{Harz:2012fz,Harz:2014gaa,Harz:2014tma,Baker:2015qna,Ellis:2015vaa,Harz:2016dql,Ellis:2015vna,Pierce:2017suq}.

Moreover, it is well established that if DM or its co-annihilating partners couple to significantly lighter force mediators, then non-perturbative effects -- the Sommerfeld enhancement of the annihilation processes~\cite{Hisano:2002fk,Hisano:2003ec} and the formation of unstable bound states~\cite{vonHarling:2014kha} -- become important. It has been shown, in particular, that the electroweak gauge boson exchange, as well as the gluon exchange in the case of DM co-annihilating with colored particles, can affect significantly the density of WIMP DM with mass as low as $\sim 500 \, \GeV$~(see e.g.~\cite{Hryczuk:2011tq,Cirelli:2015bda,Beneke:2016ync,Liew:2016hqo,Biondini:2017ufr}). On the other hand, the Higgs exchange has been neglected so far, or studied only insufficiently~\cite{LopezHonorez:2012kv}. The rationale has been twofold: The Higgs boson, being heavier, yields a shorter-range force than the SM gauge bosons, and the DM coupling to the Higgs is in many models smaller than, or only comparable to the SM gauge couplings.

In this work, we demonstrate that, contrary to the above expectation, the \emph{Higgs enhancement} can be significant. We employ a simplified model in which DM coannihilates with colored particles that couple to a SM-like Higgs. While this setup has wider applicability, it is inspired by the Minimal Supersymmetric SM (MSSM) and the measurement of the Higgs mass, which together motivate light stops with large coupling to the Higgs~\cite{Haber:1990aw,Haber:1996fp}. Related DM studies in the MSSM have been conducted recently~\cite{Keung:2017kot,Ibarra:2015nca,Pierce:2017suq}. Moreover, this setup allows to directly compare the effect of the Higgs exchange with that of other mediators, in particular the gluons. As we show, the Higgs enhancement is significant even for moderate couplings to the Higgs, and can be comparable to the gluon exchange.

This letter is organized as follows: After specifying the simplified model, we review the DM freeze-out in the presence of coannihilations. We describe the effect of the Higgs enhancement on the annihilation cross-section, before demonstrating its impact on the DM abundance. We conclude with a discussion of the Higgs enhancement in the MSSM, and its experimental implications.

\section{Simplified Model}

We assume that DM is a Majorana fermion $\chi$ of mass $\mx$, that coannihilates with a complex scalar $X$ of mass  $\mX$. $\chi$ and $X$ are the lightest and next-to-lightest particles (LP and NLP) odd under a $\mathbb{Z}_2$ symmetry that prevents the LP from decaying. For our purposes, we need not specify the $\chi$ interactions with $X$ or other particles further.  
$X$ transforms as a {\bf 3} under $SU(3)_c$, and couples to a real scalar $h$ of mass $\mh=125\,\GeV$, via
\begin{align}
{\d \cal L} &= (D_{\m,ij} X_j)^\dagger \, (D_{ij'}^\m X_{j'}) - \mX^2 \, X_j^\dagger X_j^{}  
\nn \\
&+ \frac12 (\partial_\m h) (\partial^\m h) - \frac12 \mh^2 h^2- \gh^{} \mX \ h \, X_j^\dagger X_j^{}  \,.
\label{eq:Lagrangian}
\end{align}
Here, $D_{\m,ij} = \d_{ij} \partial_\mu - i \gs \, G_\m^a T^a_{ij}$, with $G_\m^a$ being the gluon fields and $T^a$ the corresponding generators. In a complete model, the scalar potential includes also the quartic terms. Moreover, a SM-like Higgs would couple to the SM particles. For simplicity, we do not consider these couplings, whose effect is well known. We focus instead on the long-range effect of the $h \, \Xdagger_j X_j^{}$ term only.
The $h \, X_j^\dagger X_j^{}$ coupling is expressed in terms of $\mX$ for convenience. It is not necessarily proportional to $\mX$ in complete models, since we typically expect other sources for $\mX$ besides the Higgs vacuum expectation value. Indeed, for a scalar $X$ boson, a bare mass is allowed by all unitary symmetries. It may be forbidden by a non-unitary symmetry, such as supersymmetry, but appears as a soft supersymmetry-breaking term in the MSSM.

In this setup, if the \emph{relative} mass difference $\Delta \equiv (\mX - \mx)/\mx$ is small, the DM density is determined by the $\chi$ self-annihilation, the $\chi-X$ and $\chi-X^\dagger$ coannihilation and the $X-X^\dagger$ annihilation, as we now describe.

\section{Relic abundance}
The DM density for a system of (co)annihilating particles is determined by the Boltzmann equation~\cite{Edsjo:1997bg}
\beq
\frac{d\tilde{Y}}{dx} = - \sqrt{\frac{\pi}{45}} 
\, \frac{\gstareffsqrt \, M_{\mathrm{Pl}} \, m_\chi \, \<\sigmaeff \, \vrel \>}{x^2} \ 
\ (\tilde{Y}^2 - \tilde{Y}_\eq^2) \,,
\eeq
where $\tilde{Y}$ is the sum of the yields of all coannihilating species, $\tilde{Y} = \sum_i Y_i =  \sum_i n_i /s$ with $i=\chi,X,X^\dagger$.  $s \equiv (2\pi^2/45) \, \gstarS^{} \, T^3$ is the entropy density of the universe, 
$g_{*, \mathrm{eff}}$, $\gstarS^{}$ are the energy and entropy degrees of freedom, and $x\equiv\mx/T$ is the time parameter. The yields in equilibrium are
\begin{align}
Y^\eq_i = 
\frac{90}{(2\pi)^{7/2}} \ \frac{g_i}{\gstarS} \ [(1+\d_i)x]^{3/2} \ e^{-(1+\d_i)x}\,,
\label{eq:yieldeq}
\end{align}
with $\d_\x=0$, $\d_{X,X^\dagger}=\Delta$ and $\gx=2$, $g_{X,X^\dagger}=3$.  As seen from Eq.~\eqref{eq:yieldeq}, if the NLP is close in mass to the LP, its density is only mildly more suppressed, and it contributes to the DM density substantially.

The thermally-averaged effective cross-section includes all (co)annihilation processes weighted by the densities of the participating species. We  assume that the dominant contribution is the $X X^\dagger$ annihilation cross-section $\sigma_{\mathsmaller{X\Xdagger}}$, such that
\beq
\<\sigmaeff \, \vrel \> = 
\frac{2\YX^\eq \YXdagger^\eq \, \<\s_{\mathsmaller{X\Xdagger}} \, \vrel\> }{\tilde{Y}_\eq^2}\,,
\label{eq:sigma_eff}
\eeq
where $\vrel$ is the relative velocity. In our model, the dominant annihilation channels are the $s$-wave processes $X \Xdagger \to gg, \, hh$. The annihilations $X \Xdagger \to q \bar{q}, \, gh$ are $p$-wave suppressed, and we neglect them for simplicity. The cross-sections for the $s$-wave processes are
\begin{subequations}
\label{eq:sigmaANN}
\begin{align}
\label{eq:sigma_XXbarTogg}
(\sigma \vrel)_{X \Xdagger \to gg} &= 
\frac{14}{27} \frac{\pi \alpha_s^2}{\mX^2}
\times 
\( \frac27 \ S_0^{\bf [1]} + \frac57 \ S_0^{\bf [8]}  \)  ,
\\
(\sigma \vrel)_{X \Xdagger \to hh} &= 
\frac{4\pi \ah^2}{3\mX^2} \ \frac{(1-\mh^2/\mX^2)^{1/2}}{[1-\mh^2/(2\mX^2)]^2} 
\times S_0^{\bf [1]}   ,
\label{eq:sigma_XXbarTohh}
\end{align}
\end{subequations}
where $\ah \equiv \gh^2/(16 \pi)$~\cite{Petraki:2015hla,Petraki:2016cnz} and $\as \equiv \gs^2/(4\pi)$. As seen, $\sigma \vrel$ factorize into their perturbative values and the Sommerfeld factors $S_0$ that encapsulate the effect of the long-range interaction, which we discuss next.

Besides the direct annihilation processes~\eqref{eq:sigmaANN}, the formation and decay of $X-X^\dagger$ bound states may deplete DM significantly~\cite{vonHarling:2014kha,Ellis:2015vaa,Kim:2016zyy,Kim:2016kxt,Liew:2016hqo,ElHedri:2017nny,Biondini:2017ufr}. Since our focus here is to demonstrate the long-range effect of the Higgs, we shall neglect these processes, which involve considerable technicalities, and present a complete treatment elsewhere~\cite{HiggsBoundStates}. Suffice to say that bound-state effects imply an even stronger impact of the Higgs exchange on the DM density.

\section{Higgs enhancement}

If two particles couple to a light force mediator, then their long-range interaction distorts their wavepackets. This is known as the Sommerfeld effect~\cite{SakharovEffect,Sommerfeld:1931}, which enhances or suppresses the inelastic scattering at low $\vrel$, for attractive or repulsive interactions, respectively.

\begin{figure}[t!]
\includegraphics[width=0.46\textwidth]{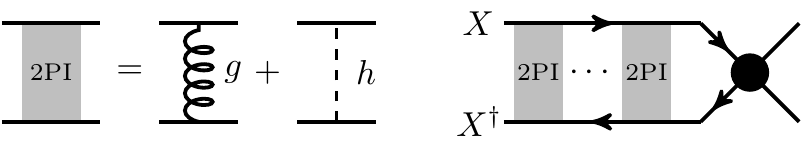}
\caption[]{\label{fig:ladder} 
The interaction of  $X,\,X^\dagger$ at infinity affects the annihilation processes. The ladder represents the resummation of the two-particle irreducible diagrams (2PI) that yield a long-range effect: the gluon and the Higgs exchange. The black blob indicates the various annihilation channels.}
\end{figure}

In our model, $X$ and $X^\dagger$ interact via the gluons and the Higgs. The long-range effect of these interactions is captured by the ladder diagrams shown in Fig.~\ref{fig:ladder}. The resummation of all two-particle irreducible diagrams amounts to solving the Schr\"odinger equation with a mixed Coulomb and Yukawa potential,  
\beq 
V(r) = - \frac{\ag}{r} - \frac{\ah}{r} \ e^{-\mh r} \,.
\label{eq:Potential}
\eeq 
The coupling $\ag$ depends on the color representation of the $X,X^\dagger$ state. The color space decomposes as ${\bf 3 \otimes \bar{3} = 1 \oplus 8}$;  in the singlet state, the gluon exchange is attractive with $\ag = (4/3) \as$, while in the octet state, the interaction is repulsive with $\ag = -\as/6$~\cite{Kats:2009bv}.

While previous works examined the gluon exchange, we consider also the attractive force mediated by the Higgs. The Higgs exchange enhances the attraction in the singlet state, and reduces, or overcomes the repulsion in the octet state. This has not been computed before, but as we show, affects the DM density significantly.

\medskip

The scattering states are described by a wavefunction $\f_{\vec k}(\vec r)$ that depends on $\vec{k} = \mu \vec{v}_{\rm rel}$, where $\mu = \mX/2$ is the reduced mass and $\vec{v}_{\rm rel}$ the relative velocity. We define the dimensionless coordinate $\vec{z} \equiv k \vec{r}$, and the parameters
\beq
\label{eq:SSparams}
\zeta_{g,h}	\equiv \frac{\mu \alpha_{g,h}}{\mu \vrel} = \frac{\alpha_{g,h}}{\vrel} , 
\qquad
d_h \equiv \frac{\mu\ah}{\mh} \,.	
\eeq
Then, $\f_{\vec k}$ is determined by the Schr\"odinger equation 
\beq
\left\{ 
\nabla_{\vec{z}}^2 +
1 +\frac{2}{z} \[ \zetag+\zetah \, \exp \(-\frac{\zetah z }{d_h} \) \]  
\right\} 
\phi_{\vec {k}}^{} = 0 \,,
\label{eq:phi diff}
\eeq
and the standard boundary condition at $r \to \infty$ of an incoming plane wave plus an outgoing spherical wave~\cite{Sakurai_QMbook}. For $s$-wave annihilation, the Sommerfeld factor
\beq
S_0 (\zetag,\zetah,d_h) \equiv |\f_{\vec{k}} (0)|^2 
\label{eq:S0 def}
\eeq
multiplies the perturbative cross-section~\cite{Cassel:2009wt}. 
In Eqs.~\eqref{eq:sigmaANN}, 
$S_0^{\bf [1]} = S_0 [4\as/(3\vrel), \, \ah/\vrel, \, \mX \ah/(2\mh)]$ and 
$S_0^{\bf [8]} = S_0 [-\as/(6\vrel), \, \ah/\vrel, \, \mX \ah/(2\mh)]$.

Due to the running of the strong coupling, the various factors of $\as$ must be evaluated at the appropriate momentum transfer $Q$. In the vertices of the $XX^\dagger \to gg$ tree-level diagram, $Q = m_X$ is the momentum of the radiated gluons. In the ladder diagrams that determine the Sommerfeld factors, $Q = \mu \vrel$ is the average momentum transfer between $X,X^\dagger$. The running of $\alpha_s$ is implemented according to~\cite{Patrignani:2016xqp,Prosperi:2006hx}. $\alpha_h$ is also subject to running that can be computed within UV complete models. Here we neglect this effect, which however is not expected to change our conclusions.

\medskip

\begin{figure}[t]
\centering
\includegraphics[width=0.46\textwidth]{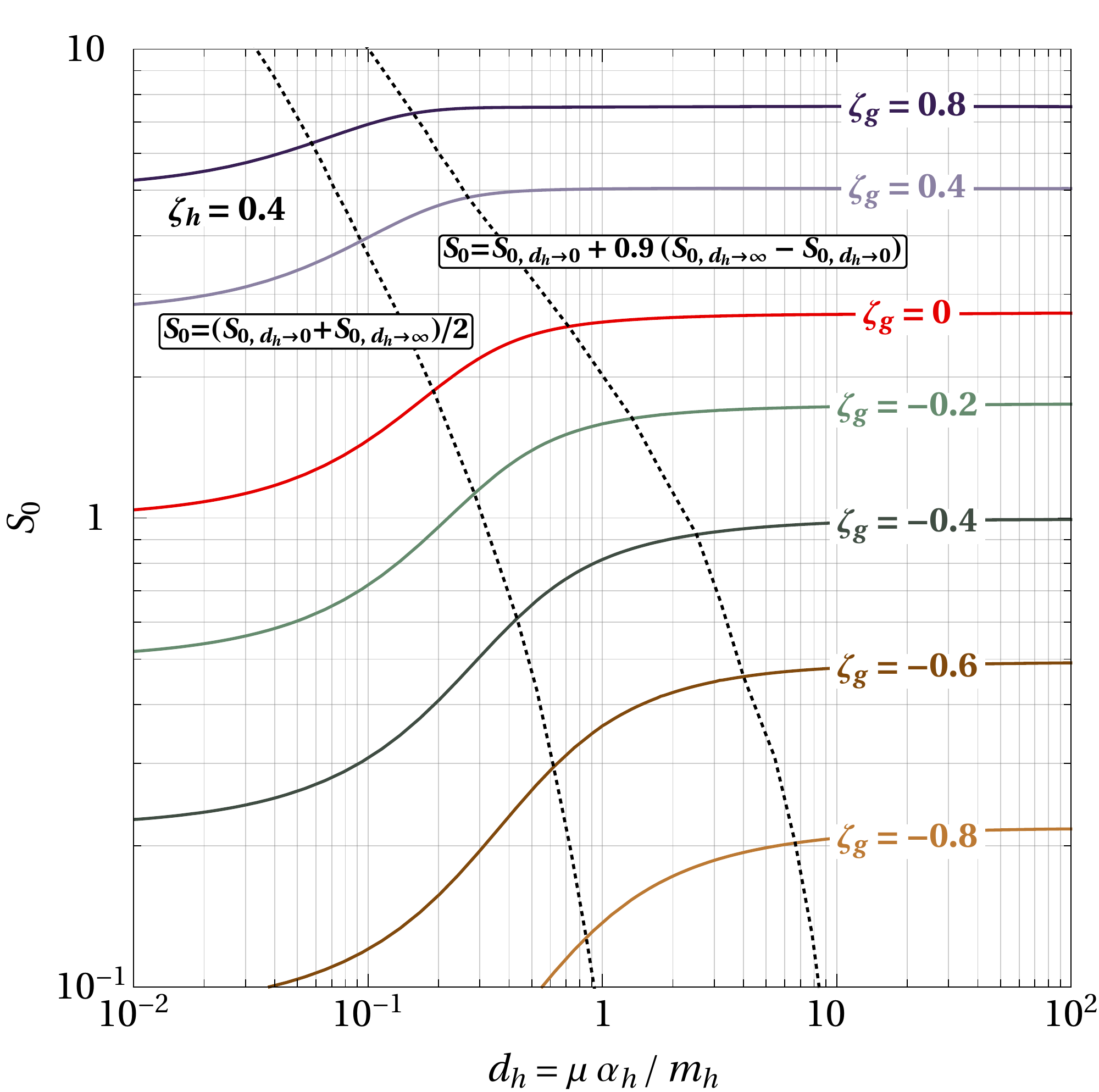}
\caption[]{\label{fig:S0vsXi}
The Sommerfeld factor in the mixed Coulomb and Yukawa potential of Eq.~\eqref{eq:Potential}. At $d_h \to 0$~and~$\infty$, the Coulomb limit is attained, $S_0 \simeq 2\pi \zeta/(1-e^{-2\pi\zeta})$, with $\zeta = \zeta_g$~and~$\zeta_g + \zeta_h$, respectively. The \emph{dotted black lines} mark the $d_h$ values for which $S_0$ is above its $d_h \to 0$ limit by 50\% (left) and 90\% (right) of the difference toward its $d_h \to \infty$ limit.
}
\end{figure}

\begin{figure}[h!]
\centering
\includegraphics[width=0.48\textwidth]{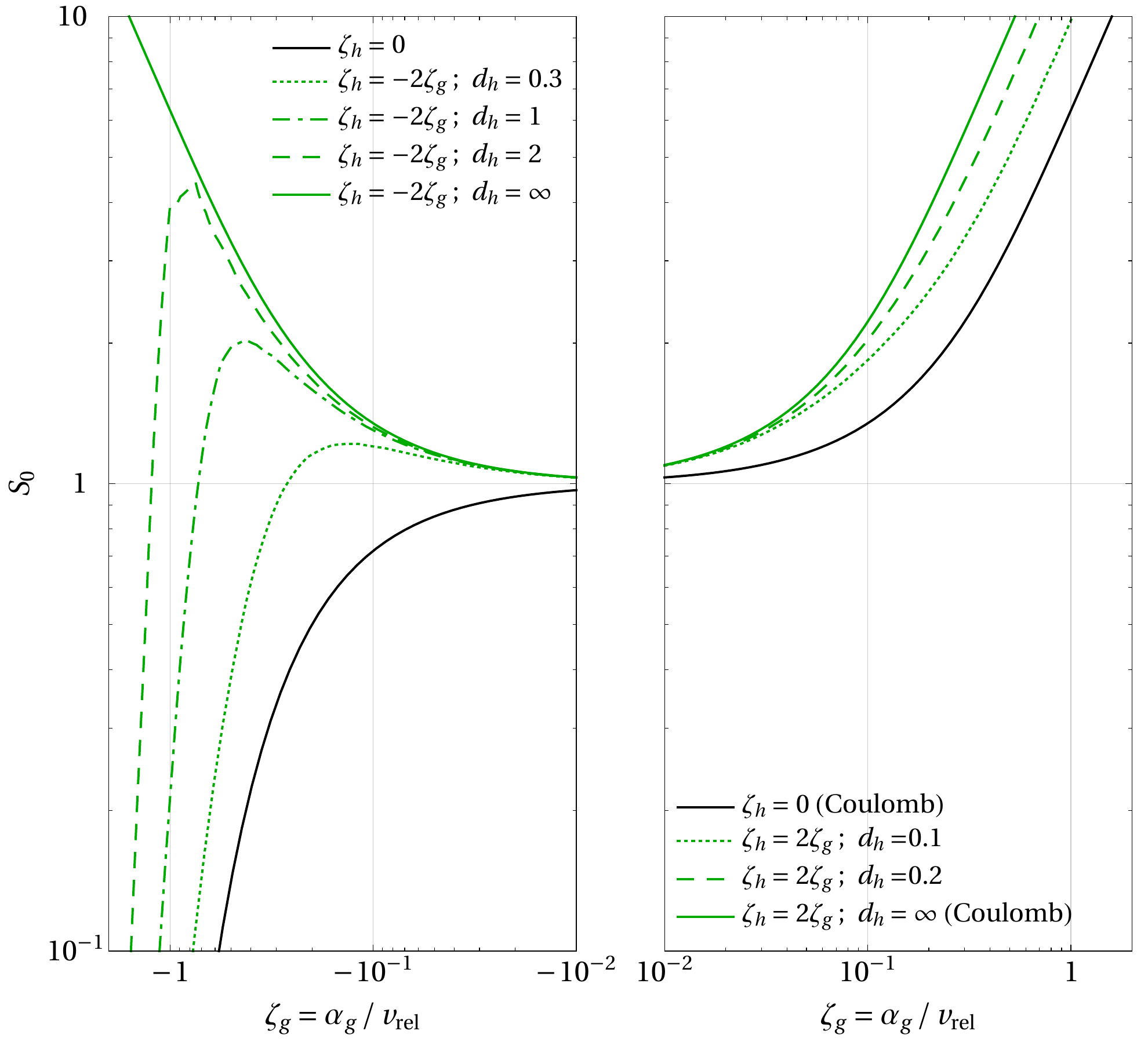}
\caption[]{\label{fig:S0vsZetag}
$S_0$ vs $\zeta_g$, for both repulsive (\emph{left panel}) and attractive (\emph{right panel}) Coulomb interaction. We take the Yukawa coupling to be twice as strong as the Coulomb one, $\alpha_h=2|\alpha_g|$, and vary the parameter $d_h \equiv \mu \alpha_h/m_h$. }
\end{figure}

The dimensionless parameters of Eq.~\eqref{eq:SSparams} contrast physical scales. The Bohr momentum $\mu \alpha$ indicates the momentum transfer around or below which non-perturbative effects arise; the parameters $\zeta_g$ and $\zetah$ compare this scale with the average momentum exchange between the interacting particles, $\mu v_{\mathrm{rel}}$. For the Yukawa term, $d_h$ compares the Bohr radius $(\mu \alpha_h)^{-1}$ with the range of the potential, $m_h^{-1}$. It is typically expected that the Sommerfeld effect arises roughly for $|\zeta_g|\gtrsim {\cal O}(1)$ and $\zeta_h, d_h \gtrsim {\cal O}(1)$, for the Coulomb and Yukawa potentials respectively, and that the Coulomb limit of the Yukawa potential is attained for $d_h\gtrsim\zeta_h$ ($\mu \vrel \gtrsim \mh$). Circumscribing this parameter range more precisely is important for anticipating the phenomenological implications of force mediators. We find that the interplay of the Coulomb and Yukawa terms affects this determination. In Figs.~\ref{fig:S0vsXi} and~\ref{fig:S0vsZetag}, we illustrate this point.


In Fig.~\ref{fig:S0vsXi}, we explore the dependence of $S_0$ on $d_h$. We fix $\zeta_h=0.4$; given the typical values of the velocity during freeze-out, $\vrel \sim 0.2$, this corresponds to sizable but moderate couplings, $\ah \sim 0.08$. The \emph{red line}, $\zeta_g=0$, shows the Higgs enhancement without the effect of gluon exchange. We may observe that for $d_h$ as low as $\sim 0.2$ the enhancement is substantial, $S_0 \sim 1.8$, and amounts to about half of the enhancement of the Coulomb limit. For the SM Higgs mass of $125\,\GeV$, $d_h \sim 0.2$ is realized for moderate $\ah$ and low enough $\mX$ to be probed by collider experiments; for instance,  $\ah = 0.1$ and $\mX=500\,\GeV$ or $\ah = 0.05$ and $\mX = 1\,\TeV$.

Let us next consider the superposition of the Yukawa and the Coulomb potentials. The \emph{dotted black lines} in Fig.~\ref{fig:S0vsXi} depict an important trend. For an attractive Coulomb term, the Yukawa contribution to $S_0$ is significant (with respect to its full capacity that is determined by its Coulomb limit) for even lower $d_h$ than in the case of a pure Yukawa potential. That is, the Yukawa interaction manifests as long-range  for an even smaller hierarchy between scales. 
This feature has not been identified in DM studies before, but has a significant implication. 
It indicates that the Higgs enhancement is important for lower $\ah$ and/or $\mX$ than anticipated.

It is also striking that the Yukawa attraction can considerably ameliorate or fully overcome the suppression due to the repulsive Coulomb potential still for modest values of $d_h$. For this to occur, it is of course important that $\ah$ is at least comparable to $|\ag|$. In the model considered here, the Coulomb coupling in the octet state is $\ag = -\as/6 \sim -0.02$, where we took $\as \sim 0.1$. It follows that even weak couplings to the Higgs can substantially enhance the annihilation rate of the octet state.

Nevertheless, the Coulomb suppression is exponential in $\zetag$, and dominates over the Yukawa attraction at low $\vrel$ (large $|\zetag|$), as seen in Fig.~\ref{fig:S0vsZetag} (left). This does not change our earlier conclusion though, since the DM relic density is determined mostly at earlier times, when $|\zetag|$ is not much larger than 1.

\section{Impact on the relic density}

\begin{figure}[t]
\centering
\includegraphics[width=0.46\textwidth]{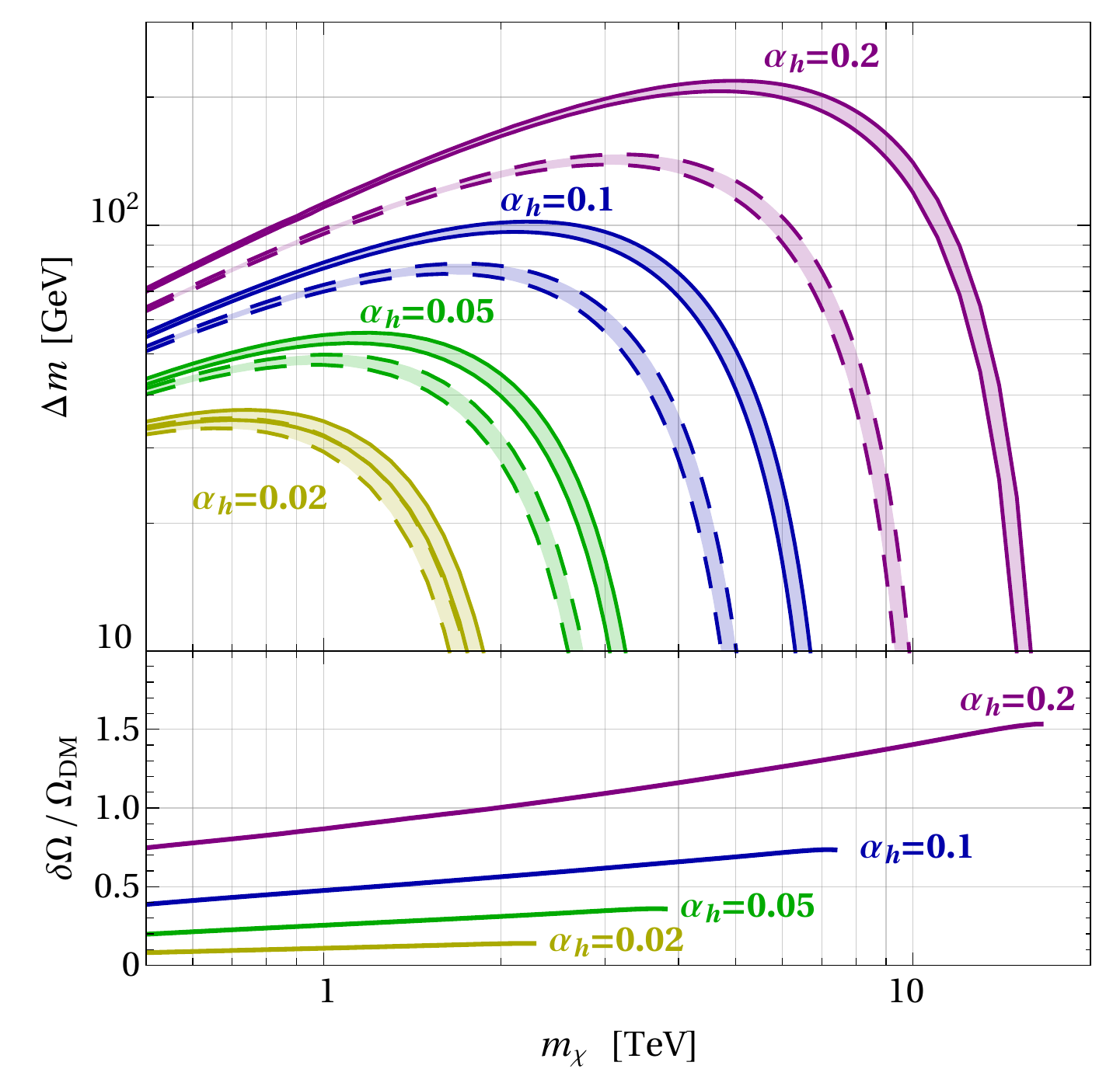}
\caption[]{\label{fig:relic}
\emph{Upper panel:} The mass difference between the LP and the NLP vs the DM mass, for different couplings to the Higgs. The bands indicate the $3 \sigma$ uncertainty on $\Omega_{\mathrm{DM}}$. 
The \emph{dashed lines} include only the effect of gluon exchange, while the \emph{solid lines} incorporate the Higgs enhancement.
\emph{Lower panel:} The effect of the Higgs enhancement on the DM density ranges from 10\% to 150\%, for the parameters considered. 
$\Delta m$ is fixed for every $m_{\rm DM}$ and $\alpha_h$, by the full freeze-out computation that includes the Higgs exchange.
}
\end{figure}

\begin{figure}[t]
\centering
\includegraphics[width=0.45\textwidth]{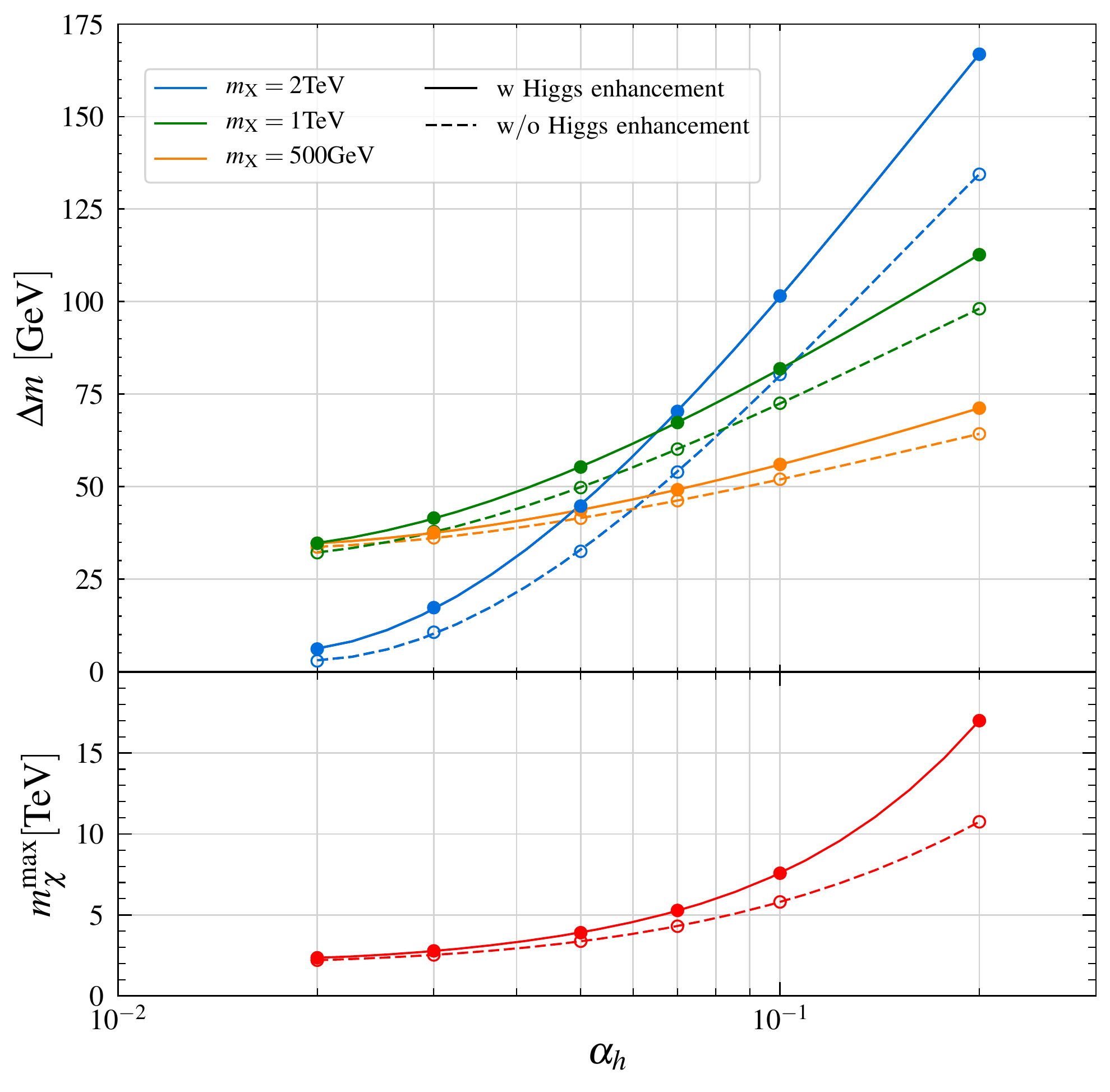}
\caption[]{\label{fig:deltam}  
The mass difference $\Delta m = m_{X}-m_{\chi}$ for $m_{\chi}=(0.5, 1, 2) \, \mathrm{TeV}$ (\emph{upper panel}), and the maximal DM mass (\emph{lower panel}) versus $\alpha_h$, with Higgs enhancement (\emph{solid lines}) and without (\emph{dashed lines}).
}
\end{figure}

The above discussion suggests that the impact of the Higgs exchange on the DM density can be significant. This is indeed so. We compute the DM freeze-out with and without the Higgs enhancement, and present our results in Fig.~\ref{fig:relic}, for $\ah$ in the range $0.02 - 0.2$.\footnote{
We may estimate the value of $\ah$ around which our computation breaks down by
considering the upper bound on the inelastic cross-sections implied by the $S$-matrix unitarity~\cite{Griest:1989wd}. Applying the $s$-wave unitarity bound on Eq.~\eqref{eq:sigma_XXbarTohh}, we find $\ah<0.7$ (neglecting $\as$). Around and above this value, higher order corrections 
must be considered (see Refs.~\cite{vonHarling:2014kha,Baldes:2017gzw} for related discussion). Note that this condition is stronger that the commonly assumed perturbativity condition $\ah < 4\pi$.
Depending on the UV completion of the theory, additional unitarity and perturbativity conditions may be pertinent. 
} 
The enhanced annihilation rate implies that $\Omega_{\DM}$ is obtained for larger $\Delta m$ and/or $m_{\chi}$, as shown in the upper panel. Already for $\ah = 0.02$, the effect exceeds the $3\sigma$ experimental uncertainty in $\Omega_{\DM}$, while for larger $\ah$ it becomes very severe, as seen in the lower panel.
While here we focus on $\mX \geqslant 500$~GeV to be on par with experimental constraints, the Higgs enhancement can affect the DM density significantly even for lower masses, as can be extrapolated from the lower panel of Fig.~\ref{fig:relic}, and deduced from Figs.~\ref{fig:S0vsXi} and \ref{fig:S0vsZetag}, by appropriate estimates.

In the upper panel of Fig.~\ref{fig:deltam}, we focus on low  DM masses, $m_{\chi}=$~(0.5, 1, 2)~TeV that can be probed at current and future colliders, and show the considerable increase in the predicted mass gap $\Delta m$. The Higgs enhancement implies a stronger lower limit on $\Delta m$ for a given $m_\chi$ and $\ah$; additional (co)annihilation channels -- potentially also Higgs-enhanced -- that are expected in complete models, shift the prediction for $\Delta m$ to higher values.

The maximum DM mass $m_{\chi}^\mathrm{max}$ in agreement with $\Omega_{\DM}$ is shown in the lower panel of Fig.~\ref{fig:deltam}. The Higgs enhancement shifts $m_{\chi}^\mathrm{max}$ substantially, by about 150~GeV for $\ah=0.02$, to more than 6~TeV for $\ah=0.2$. 

The mass gap and the maximum DM mass have important experimental implications, as we discuss below.

\section{Higgs enhancement in the MSSM}

Sizable couplings to the Higgs occur in the MSSM, especially in light stop scenarios with maximal mixing between stop mass eigenstates, that are motivated by the value of the Higgs mass~\cite{Haber:1990aw,Harz:2012fz}. In the MSSM11, we have identified an example scenario of neutralino LP and stop NLP, with masses $m_{\tilde{\chi}_1^0}=982.5\,\GeV$ and $m_{\tilde{t}_1}=1066.1\,\GeV$, where the coupling to the Higgs amounts to $\alpha_h\simeq0.15$.  The MSSM parameters are (dimensionful quantities in GeV)  
$\tan\beta = 16.3$, 
$\mu = 2653.1$, 
$m_{A^0} = 1917.9$, 
$M_1 = 972.1$, 
$M_2 = 1944.1$, 
$M_3 = 5832.4$, 
$M_{\tilde{q}_{1,2}} = 3054.3$, 
$M_{\tilde{q}_{3}} = 2143.7$, 
$M_{\tilde{\ell}} = 2248.3$, 
$M_{\tilde{u}_{3}} = 2143.7$, 
and 
$A_t = -4380.93$. As large trilinear couplings are known to give rise to color-breaking minima in the scalar potential that could endanger the stability of the $SU(3)_c$-symmetric vacuum, we have checked with \texttt{Vevacious}~\cite{Camargo-Molina:2013qva,Camargo-Molina:2014pwa} for stability. We found that in this scenario, the color-symmetric vacuum is metastable but sufficiently long-lived. Thermal corrections imply an upper limit on the reheating temperature of $T\approx M_{\mathrm{SUSY}}$. In scenario B of Ref.~\cite{Harz:2014gaa}, with $m_{\tilde{\chi}_1^0}=1306.3~\mathrm{GeV}$ and $m_{\tilde{t}_1}=1363.0\,\GeV$, we find $\alpha_h = 0.03$. Moreover, in the parameter space recently explored in Ref.~\cite{Pierce:2017suq}, we estimate $\ah$ to be in the range $\sim (0.02 - 0.07)$ for $m_{\tilde{t}_1}=500~\mathrm{GeV}$ and $\sim (0.01 - 0.05)$ for $m_{\tilde{t}_1}=1500~\mathrm{GeV}$. Larger couplings to the Higgs than those mentioned above may be viable, however a proper study of the vacuum stability has to be pursued for specific scenarios. Furthermore, it has been argued that large couplings of colored particles to the Higgs may lead to a new phase in the MSSM, where the standard treatment does not apply~\cite{Giudice:1998dj}.

Clearly, substantial couplings to the Higgs appear in realistic models. While the Higgs enhancement has been neglected in previous studies, it is essential for interpreting the experimental results correctly.

\section{Further experimental implications}

Small mass gaps $\Delta m$ that are a feature of the scenarios considered here, lead to the production of soft jets from the decay of the NLP that are difficult to probe at the LHC. The Higgs enhancement implies larger mass gaps for a given DM mass (cf.~Fig.~\ref{fig:deltam}), therefore harder jets that may pass the detection threshold. This is important for mono-/multi-jet plus missing energy searches. Higgs-enhanced bound-state processes increase $\Delta m$ further~\cite{HiggsBoundStates}. Moreover, the Higgs enhancement implies that WIMP DM may be heavier than anticipated, and motivates indirect searches in the multi-TeV regime.\footnote{Indirect signals arise from the DM self-annihilations only. In our analysis, we assumed  for simplicity that they are negligible with respect to co-annihilations and/or self-annihilations of the DM co-annihilating partner. However, self-annihilations are expected to occur in complete models, and yield significant indirect signals.} 
Of course, to fully assess the experimental implications of the Higgs enhancement, proper analyses of realistic scenarios are necessary.\footnote{
An even stronger coupling of colored particles to the Higgs than considered here, may have implications for the electroweak symmetry breaking~\cite{Cornwall:2012ea,Pearce:2013yja} and colliders~\cite{Cornwall:2012ea,Pearce:2013yja,Kang:2016wqi}.}

\section{Conclusions}
We have demonstrated that the Higgs enhancement can affect the DM density significantly, thereby altering the interpretation of the experimental results within specific theories, but also motivating extended searches. Importantly, the interplay between the Higgs and other mediators -- within or beyond the SM -- affects the efficiency of the Higgs enhancement. Besides the scenarios considered here, we expect that the Higgs enhancement is important in a variety of models around or above the TeV scale, including WIMP scenarios, such as the inert doublet model~\cite{Barbieri:2006dq,LopezHonorez:2006gr,Hambye:2009pw,Goudelis:2013uca,Eiteneuer:2017hoh,Biondini:2017ufr}, as well as Higgs portal models (see~\cite{Assamagan:2016azc} and references therein).

\section*{Acknowledgements}

We thank Florian Staub for providing help with \texttt{Vevacious}, and Andreas Goudelis for useful discussions. J.H was supported by the Labex ILP (reference ANR-10-LABX-63) part of the Idex SUPER, and received financial state aid managed by the Agence Nationale de la Recherche, as part of the programme Investissements d'avenir under the reference ANR-11-IDEX-0004-02. K.P. was supported by the ANR ACHN 2015 grant (``TheIntricateDark" project), and by the NWO Vidi grant ``Self-interacting asymmetric dark matter".


\bibliography{Bibliography.bib}

\end{document}